# Trip Table Estimation and Prediction for Dynamic Traffic Assignment Applications

Sajjad Shafiei[*1], Adriana-Simona Mihăiţă[1,2], Chen Cai[1]

[1]Advanced Data Analytics in Transport, DATA61|CSIRO, Sydney, Australia
[2]University of Technology, Sydney, Australia
[*]sajjad.shafiei@data61.csiro.au

**Abstract:** The study focuses on estimating and predicting time-varying origin-destination (OD) trip tables for a dynamic traffic assignment (DTA) model. A bi-level optimisation problem is formulated and solved to estimate OD flows from pre-existent demand matrix and historical traffic flow counts. The estimated demand is then considered as an input for a time series OD demand prediction model to support the DTA model for short-term traffic condition forecasting. Results show a high capability of the proposed OD demand estimation method to reduce the DTA model's error through an iterative solution algorithm. Moreover, the applicability of the OD demand prediction approach is investigated for an incident analysis application for a major corridor in Sydney, Australia.

## 1. Introduction

Traffic forecasting is a necessary step for efficient network operation and is an integral part of intelligent transportation systems (ITS) applications. In the transport domain, dynamic traffic assignment (DTA) models are known as a reliable tool to replicate complex traffic conditions [1]. These models are mainly built based on game theory principles. Each traveller attempts to minimize his/her travel cost (time) while their decision impacts the traffic condition other traveller's decisions to move in the network [2]. By considering this essential principle, the intricate traveller route decision can be modelled in the network. DTA models are categorized into broad groups of analytical- and simulation-based models. Among different types of analytical models, variational inequality formulations are more popular because they can consider both optimization conditions and equilibrium in one formulation [3]. Restrictive assumptions often characterize the analytical approach, and they hardly represent the complicated interaction between users and the network. Also, increasing the computational power leads to a growing interest in the development of the simulation-based DTA models in recent years.

The most crucial input for any DTA models is the origin-destination (OD) trip table. The success of the DTA application relies on the quality of this fundamental input and how well it captures the movement in the city from one time-interval to another. As a result, the estimation of OD trip tables still remains a hot research topic for many years [4], [5]. Estimating the OD demand by using link traffic data is a popular approach and far superior to doing the conventional travel surveys which are slow and expensive. The advent of the new types of OD demand data such as mobile data or vehicle trajectories in recent years provides cheaper and faster methods to obtain more accurate OD demand information. However, in practice, this type of data also needs further adjustments for being usable in DTA models [6]. Many studies proposed a bi-level optimization formulation where the feedback of demand changes is evaluated by an assignment model iteratively [7]. The demand is estimated with the objective that the error between the simulated and the observed traffic is minimised. The initial studies in OD estimation considered a steady state for the transport system, and the OD demand was assumed constant during the modelling interval [5]. By increasing the congestion in the network and improving the assignment methodologies, the modelling windows are extended to a few hours or even an entire day. Under this scenario, assuming an invariant demand for such a long-time interval was an incorrect assumption. As a result, the static models are replaced with dynamic versions which can handle the variation in demand during the study time-interval. Some challenges and opportunities of dynamic OD estimation problems are being broadly studied in the literature such as the consideration of advanced traffic surveillance data [4], methodological improvements [8]–[11] and simultaneous calibration of supply and demand parameters [12], [13].

The above research efforts mostly conducted as offline demand estimation models. However, parallel research studies focus on the online estimation/prediction of OD demand matrices. Offline models usually provide an initial OD demand for online applications which can then remove the noise from the measurement data and adjust the model's parameters within a reasonable computational time. ARIMA-family models are widely employed for demand prediction [14]. Such models predict the OD demand for one or more discrete time intervals by encountering the history demand variation. The models consist of an autoregressive (AR) and a moving average (MA) parts. The principle assumption of the ARIMA models is the stationarity of the forecasted variable. In other words, the mean and variance of the variable should remain constant. However, in case of detecting a specific trend in the series, some techniques such as logarithmic transformation or differentiation can be employed to make series stationarity.

In this study, we first estimate the dynamic OD trip tables through an optimization framework. Thereafter, OD flows are predicted for the next time-intervals by using the traditional ARIMA models. Although it has been shown that the performance of some more advanced OD prediction methods can outperform the ARIMA, its prediction is usually considered as a reliable baseline. Finally, the calibrated DTA model is employed to investigate traffic prediction in an incident analysis case study for a major corridor in Sydney, Australia. The rest of the paper is organized as follow. The next section describes the methodology proposed in this study. Section 3 presents some results for the application of the model in incident analysis. The last section provides a conclusion and draws some future research extensions.

## 2. Methodology

The essential objective of the OD demand estimation problem is to minimize the error between the simulated and the observed traffic measurements. The OD demand estimation problem is expressed as a system of equations in which the unknown parameters are OD flows, and each equation represents the combination of OD flows crossing an observed traffic link. Once the system of equations is solved, a set of estimated OD demand flows is obtained which should be evaluated by using a traffic assignment model. If simulated flows are close enough to the observed flows, the estimated OD demand matrices are accepted as a final solution. Otherwise, the mentioned procedure should be repeated until the desired termination criterion is met. In this study, we used a bi-level optimization problem in which the upper level contains the system of equations to be solved and a DTA model in the lower level evaluates the changes in the OD demand. In addition to reducing the deviation between simulated and observed traffic data, another term in the upper level seeks to keep the estimated demand as close as possible to the initial demand. The latter term assists the solution to avoid merging to the local solutions which far from the initial demand flows. We express the problem mathematically as follows:

$$\min \omega . \sum_{i \in I} \sum_{t=1}^{T}(x_i^t - \hat{x}_i^t)^2 + (1-\omega). \sum_{a \in A} \sum_{t=1}^{T}(y_a^t - \hat{y}_a^t)^2 \quad (1)$$
$$y_a^h = \sum_{i \in I} \sum_{t=1}^{h} p_{a,i}^{h,t}(X) x_i^t$$

where,
$\hat{x}_i^t, x_i^t$ are the initial and estimated demand flow of OD pair $i$ ($i \in I$) at time period $t$,
$\hat{y}_a^h, y_a^h$ are the observed and estimated link flow in link "$a$" at a time period $h$ ($a \in A$, $h=[1,T]$),
$p_{a,i}^{h,t}$ is the assignment proportion of $x_i^t$ that passes link "$a$" during a time period $h$,
$\omega$ is the reliability weight on the initial demand data.

We implemented the Problem (1) in the GAMS platform and used the KNITRO solver [15] to obtain the solution [16]. The termination criterion is assumed to be met when the variation of the model's fit measurement is less than a specific value in successive iterations. R-squared ($R^2$) is selected as the goodness-of-fit measurement for N observed link at T time intervals. Equation 2 calculates R-squared as:

$$R^2 = 1 - \frac{\sum_{a \in A} \sum_{h=1}^{T}(\hat{y}_a^h - y_a^h)^2}{\sum_{a \in A} \sum_{h=1}^{T}(\bar{y}_a^h - y_a^h)^2}, \bar{y}_a^h = \frac{\sum_{a \in A} \sum_{h=1}^{T} y_a^h}{N \times T} \quad (2)$$

The estimated demand for Problem (1) is considered a training demand set for the demand prediction module. In this study, we use ARIMA technique to forecast OD demand.

Two ARIMA's characteristics make the application of the model desirable for demand prediction models. First, the ARIMA reggresses demand fluctuation with the lagged demand values. Therefore, demand flows can be predicted when the other exogenous variables are unavailable in real-time. Second, the ARIMA essentially inclines to concentrate on the means and it less digress to the extremes. This model's behaviour suits for prediction of OD flow which usually has a smooth trend. In contrast, for link flow prediction due to the transition from free-flow conditions to stop-and-go traffic state, the perdition should shift quickly to extreme values [14]. Given mentioned reasons, the OD demand flows are predicted through ARIMA models based on the 4-hour morning peak period from 6:00 AM to 10:00 AM. In general, the ARIMA model with "$p$ autoregressive terms" and "$q$ moving-average terms" takes the following form:

$$x_i^t = \sum_{l=1}^{p} \varphi^l x_i^{t-l} + \sum_{l=1}^{q} \theta^l \epsilon^{t-l} + c + \epsilon^t \quad (3)$$

Producing an ARIMA model requires defining $p$ and $q$ in order to specify formulation. Identification of $p$ and $q$ terms involves investigating a tentative formulation for the model as a starting point. $p$ and $q$ are initially suggested to be taken from patterns either in the autocorrelation function and/or partial autocorrelation function of the series itself, or from the residuals of a previously estimated model [17]. After the general model is specified, the parameters $\varphi^l$, $\theta^l$ and $c$ are estimated through a least-squares approach.

## 3. Case Study

The large-scale Sydney transport model is implemented at three levels of macroscopic, mesoscopic and microscopic modelling and includes both private and public transport classes [18]. The large-scale macroscopic model is divided into several mesoscopic and microscopic subnetworks. The macroscopic large-scale Sydney metropolitan network includes 72,065 sections, 41,063 detectors, almost 3,600 public transport lines and 2,348 traffic zones. The model is mainly used for the long-term planning purposes and to provide any transversal demand extraction or baseline road network indications for several parallel sub-networks. For any desired sub-network to be used for mesoscopic or microscopic traffic modelling, the model can generate the corresponding initial traffic demand which will be further refined through calibration and used for prediction purposes.

The sub-networks are often modelled at the mesoscopic simulation level, however, in a case where more traffic details are required the microscopic simulation can be easily deployed as well. Although deployment of such hierarchical transport model is being acknowledged as an efficient approach to deal with the intensive computational burden of a large-scale network [19], it needs different calibration methods for each simulation level. In this study, we evaluate the proposed demand estimation and prediction models for one of the major subnetworks, the Victoria road corridor. The initial demand used in this study was obtained from the Sydney Transport Model (STM). The spatial configuration of the large-scale Sydney transport network and Victoria corridor are presented in Figure 1 a) and b) respectively.

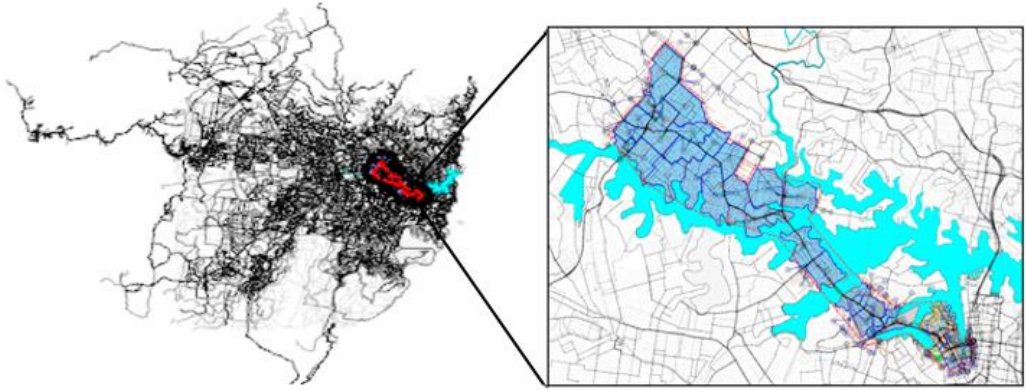

**Fig 1.** (a) Large-scale Sydney metropolitan network and (b) Victoria corridor sub-network.

We implemented the proposed demand estimation/prediction model on the Victoria corridor. The simulation is conducted for a 4h morning peak hours from 6:00-10:00 AM using initial OD matrices extracted from the large-scale model and further calibrated against available traffic flow counts for a regular weekday in March 2017. The only parameter which is predefined in Problem (1) is the reliability weight (ω) on the deviation functions of demand and link flows. Considering a high value for this parameter means that the initial demand is reliable and the estimated demand keeps close to the initial demand. In contrast, low values indicate that we do not trust the initial OD flows and the estimated OD flows can be significantly different from the initial ones. In general, it is desirable to have more reliable initial OD flows and force the solution to pick up OD flows around the initial values. Otherwise, the solution of problem (1) can result in some inaccurate OD flows which make some unrealistic traffic conditions. Figure 2 compares the objective function convergence with respect to different ω values. Selecting very small values of the reliability weight (see the red line) caused sudden changes in the OD demand values and consequently determined the appearance of gridlock in the network. However, as can be seen, choosing a very high value for the reliability weight (ω = 0.99), reduce the efficiency of the solution algorithm to minimise the objective function. The best result obtained when ω equals to 0.9. The solution smoothly converges to the optimum points and the objective function value significantly reduces from 700,000 to around 200,000. Defining an appropriate reliability weight can change from case to case and it is often selected through a trial and error process [19].

Figure 3 presents the scatter plots of observed link volumes versus simulated link volumes in four consecutive one-hour time-intervals before and after the implementation of the demand estimation process.

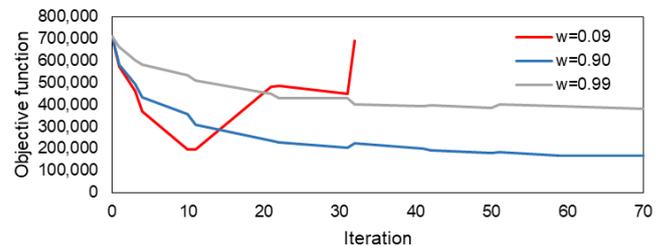

**Fig 2.** Convergence of the objective function across the number of iterations.

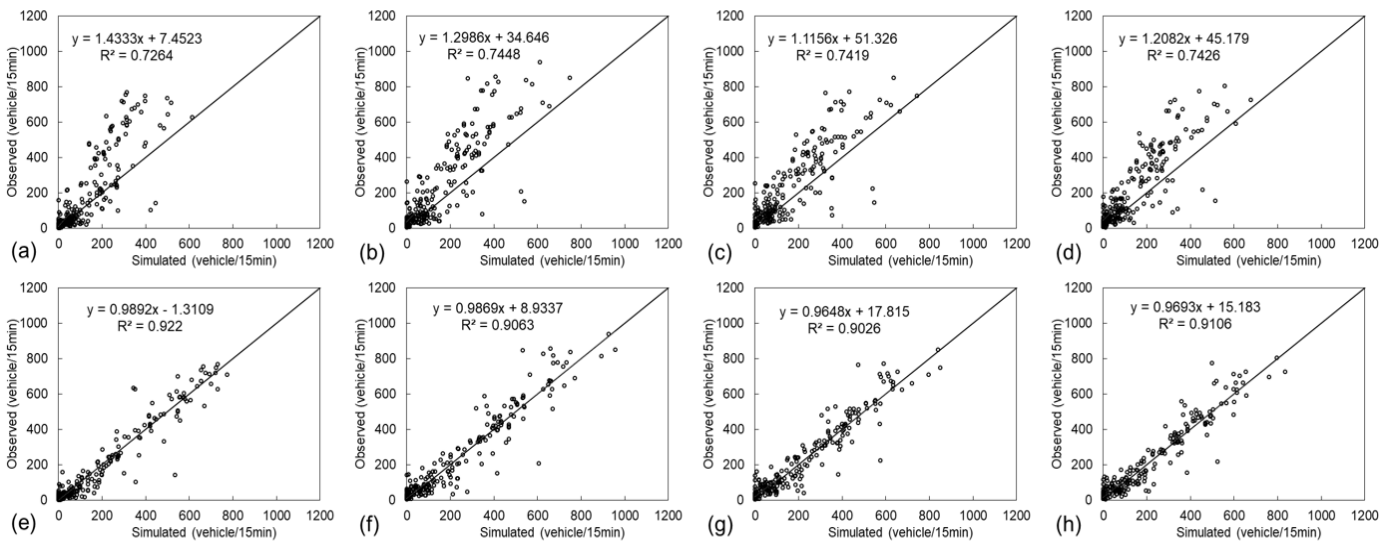

**Fig 3.** Simulated versus observed traffic volumes in four consecutive one-hour interval; before OD estimation implementation (a) 6:00–7:00 am, (b) 7:00–8:00 am, (c) 8:00–9:00 am, and (d) 9:00–10:00 am. After OD estimation implementation (e) 6:00–7:00 am, (f) 7:00–8:00 am, (g) 8:00–9:00 am, and (h) 9:00–10:00 am.

The calibrated 4h OD demand for the Victoria corridor was then transferred to the demand prediction model. There are over 3,000 OD pairs with various demand profiles in the study network. For each OD pairs an ARIMA model should be fit. Different parameters need to be specified (p, i, q) in the ARIMA model. Defining different models' specification for each OD pairs requires a vast amount of modelling effort which would be extremely computationally intensive. As a result, we consider a unique model specification for all OD demands and estimate the model's parameters ($\varphi^l$, $\theta^l$ and $c$) for each OD pairs. To identify the best model specification (p, i, q), we test a few combination of lagged values on the OD demand series. We consider the last 30 minutes as a validation time window and employed two measures of goodness. Due to the high number of small demand values in the series, we use normalized root mean squared error (NRMSE) in addition to the classical R-squared value to better compare the prediction model's performance. The NRMSE is calculated as

$$\text{NRMSE} = \sqrt{\frac{\sum_{a \in A} \sum_{t=T-1}^{T} (x_i^t - \hat{x}_i^t)^2}{\sum_{a \in A} \sum_{t=T-1}^{T} (x_i^t + \hat{x}_i^t)}} \quad (4)$$

We first applied a naïve approach to forecast the demand model ($x_i^{t+1} = x_i^t$) meaning that the demand value of the last time interval will be the same for the next 30 minutes with no major changes. This represent a baseline for further OD prediction approaches. Thereafter, four other ARIMA models' specifications are compared with the test data set. Results obtained from different models are compared in Figure 4.

As observed, all the estimated models have a better performance compared to the naïve estimation approach and among them, the ARIMA (1,0,0) (AR-1) and ARIMA (0,0,1) (MA-1) have the smallest errors that can be used for demand prediction (see Figure 5b and 5d)). Considering the demand values at 9:30 AM up to 10:00 AM led to a NRMSE=0.97 while by implementing these two prediction models the error reduced to NRMSE=0.67 meaning a 30% improvement. Therefore, with an accuracy of 0.939, we are capable of predicting accurately the OD matrices for the next half an hour. These matrices can then be used in the meso/micro simulation model in order to obtain the predicted travel times in the road network.

## 4. Incident Management Application

In this section, we employ the proposed OD demand estimation and prediction for an incident analysis application. As earlier mentioned, we use the AIMSUN mesoscopic model for the Victoria subnetwork due to the reasonable computational time and acceptable replication of the traffic dynamics in the network. This model was built based on simplified car-following and lane changing model [20]. In this model, the acceleration and deceleration of vehicles assume infinity and this simplification reduces the computational burden significantly [21]. However, the simplified model fails to fully capture the complex impacts of vehicles' lane changes caused by the incident. To address this issue, we used the microscopic simulation by selecting a 2 kilometres area surrounding the location of the incident. It is notable that once an incident was detected and the corresponding information is transferred to the DTA model, the boundaries of the microsimulation model can also be automatically determined. This process has been tested on a personal laptop Intel core i7 which took less than 30s.

We then considered a reported incident on the Anzac Bridge towards the Sydney central business district (CBD) as shown in Figure 5. Two lanes have been affected by the incident which took place at 09:57 AM. The incident information (location and the number of lanes affected) are transferred to the simulation model automatically from the incident detection model [18]. On the other side, the demand prediction module is triggered to forecast the demand starting from 10:00 AM for the next half hour. We used the AR-1 model due to its high performance previously proved in model validation step. Since the incident is located in a crucial link, there are not too many routes that travellers can take to avoid the incident. Therefore, we explore the impact of the incident duration on the total travel time in the network and the extra delay it caused in the next half an hour.

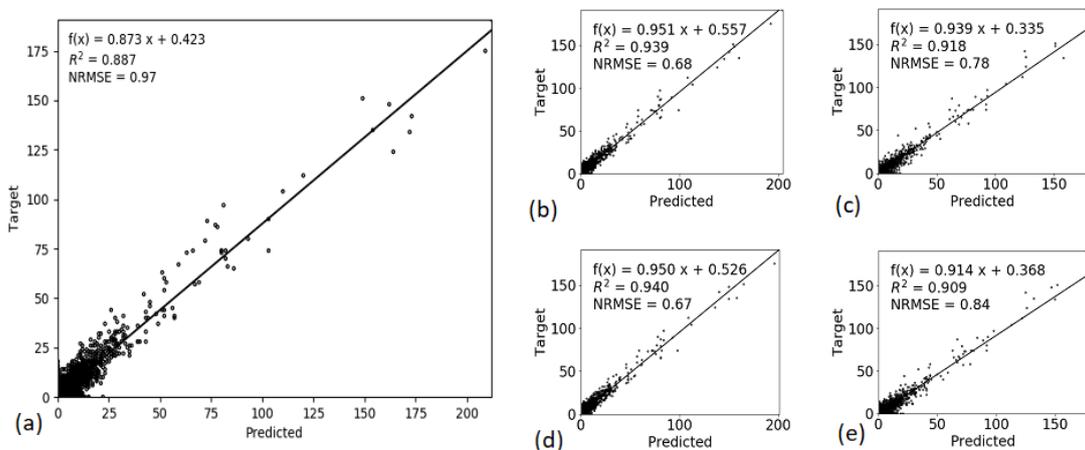

**Fig 4.** Validation of demand forecasting models: (a) no model, ARIMA (b) *(1,0,0)*, (c) *(1,1,0)*, (d) *(0,0,1)*, (e) *(0,1,1)*.

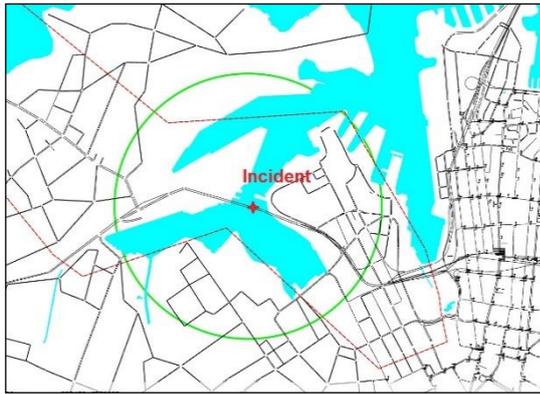

**Fig 5.** The configuration of the incident area, Anzac Bridge, Sydney.

Figure 6 presents the impact of an incident duration on the traffic flows and network delay. We basically simulated what would be the impact on the traffic flow if the incident would last for 3minutes, 5 minutes, 7 minutes and 10 minutes. Figure 6 shows that a short incident duration (less than 5 minutes) have not significantly reduced the link performance in the next 30 minutes time-interval. However, if the incident would last more than 7 or even 10 minutes, the impact on the flow is significantly higher (which is reducing from almost 10,000 to 8,000 veh/hr – a 20% reduction in the traffic flow in the morning peak hours). That means that severe bottleneck will appear in the upper stream of the network and the impact on the network will be more sever and it will take a longer time before the traffic will re-establish to a normal condition.

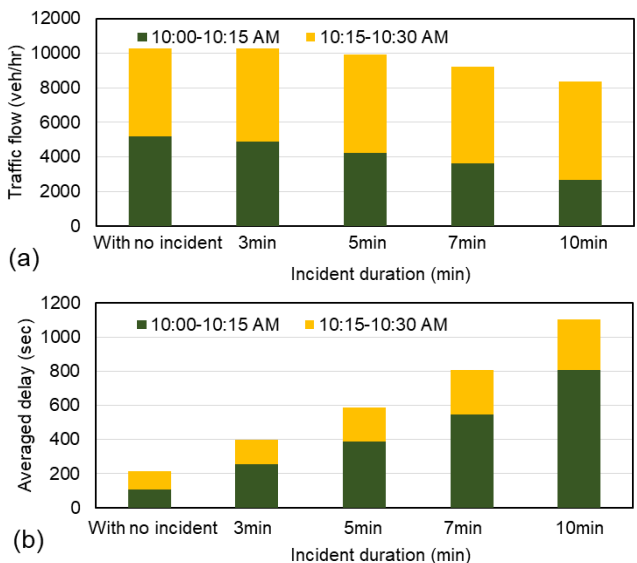

**Fig 6.** (a) Traffic flow passing through the bridge (b) Averaged delay for travellers in the area.

Figure 6 showcases once again the average delay on the network which re-iterates the previous finding: every 3 minutes increase in the total duration of an incident would lead in time to an almost exponential increase of the experienced network delay. For example, an incident of 10 minutes would produce an average delay which is almost 5 times higher than if the network wouldn't experience any incident at all.

The finding showcases the power of the proposed approach and the benefits that it brings to traffic management centres for which every minute in in the incident clearance is extremely importance. By coupling together OD prediction and microsimulation models one can accurately estimate the total impact of real accidents and their effect on the network.

## 5. Conclusion

This study proposed an efficient OD estimation and prediction approach to reinforce a simulation-based DTA model for operational and planning applications. The proposed approach initially calibrates the DTA model based on the most updated archived traffic data and then use it for demand prediction for the next time interval. A bi-level dynamic OD demand estimation problem was formulated and solved iteratively. The performance of the model was also investigated for the short-term OD prediction. Results show the high capability of the proposed dynamic OD demand estimation to improve the goodness of fit. Moreover, an application of the well-calibrated model ($R2 = 0.93$) is presented to show the applicability of the model for daily network operation purposes under incident circumstances. Worthing to note that this paper represent an ongoing study for incident analysis in real-time applications. The authors have to further validate the proposed approach using more real incidents data to ensure the applicability of the proposed framework in real-world incident management analysis.

*Limitations, future extensions*

Several limitations exist in this study that can be addressed for future studies. Given the available traffic data, we only considered a 4-hour OD demand as the training time interval. It is preferable to extend this time interval to entire day. Moreover, advanced machine learning techniques can be used to investigate their performances with the classical time series approach proposed in this paper. Finally, it is well-acknowledged that calibrating the model based on traffic volumes alone may be inefficient particularly in highly congested urban areas. Integrating other traffic information can increase the reliability of the proposed calibration framework for real applications.

## 6. Acknowledgments

The work presented in this paper is partially funded by the New South Wales Premiere's Innovation Initiative. The authors of this work are grateful for the work and support of the Traffic Management Centre from Transport for New South Wales Australia. Data61 is funded by Australian Federal Government through Commonwealth Scientific and Industrial Research Organization.